\documentclass[usenatbib,onecolumn]{mn2e}
\usepackage[pdfpagemode={UseOutlines},bookmarks,bookmarksopen,colorlinks,linkcolor={black},citecolor={black},urlcolor={black},breaklinks]{hyperref}
\usepackage{natbibmnfix}
\usepackage{astrojournals}
\usepackage{epsfig}
\usepackage{amssymb}
\usepackage{multirow}
\usepackage{multicol}
\usepackage{color}
\usepackage[varg]{txfonts}
\usepackage{url}
\usepackage{lastpage}

\date{}
\pagerange{\pageref{firstpage}--\pageref{LastPage}}
\pubyear{2014}

\begin{document}
\label{firstpage}

\title[Retrograde circumbinary resonances]{Resonances in retrograde circumbinary discs}
\author[Nixon \& Lubow]
{Chris~Nixon$^{1}$\thanks{chris.nixon@jila.colorado.edu}\thanks{Einstein Fellow}
\&
Stephen~H.~Lubow$^2$
\vspace{0.1in}\\ 
$^1$ JILA, University of Colorado \& NIST, Boulder CO 80309-0440, USA\\
$^2$ Space Telescope Science Institute, 3700 San Martin Drive, Baltimore, MD 21218, USA
} 
\maketitle

\begin{abstract}
We analyse the interaction of an eccentric binary with a circular coplanar circumbinary disc that rotates in a retrograde sense with respect to the binary. In the circular binary case, no Lindblad resonances lie within the disc and no Lindblad resonant torques are produced, as was previously known. By analytic means, we show that when the binary orbit is eccentric, there exist components of the gravitational potential of the binary which rotate in a retrograde sense to the binary orbit and so rotate progradely with respect to this disc, allowing a resonant interaction to occur between the binary and the disc. The resulting resonant torques  distinctly alter the disc response from the circular binary case. We describe results of three-dimensional hydrodynamic simulations to explore this effect and categorise the response of the disc in terms of modes whose strengths vary as a function of binary mass ratio and eccentricity. These mode strengths are weak  compared to the largest mode strengths expected in the prograde case where the binary and disc rotate in the same sense. However, for sufficiently high binary eccentricity, resonant torques open a gap in a retrograde circumbinary disc, while permitting gas inflow on to the binary via gas streams. The inflow results in a time varying accretion rate on to the binary that is modulated  over the binary orbital period, as was previously found to occur in the prograde case.
\end{abstract}


\section{Introduction}
\label{intro}
We investigate the behaviour of retrograde circumbinary discs for eccentric orbit binaries. When the binary orbit is circular, retrograde circumbinary discs can behave fundamentally differently to their prograde counterparts, due to the absence of orbital resonances (\citealt{Nixonetal2011a}). In the prograde binary case, resonances between the binary and disc  transfer angular momentum from the binary to the disc. This transfer causes the binary orbit to shrink somewhat, perhaps also changing the binary eccentricity, while the disc gains angular momentum and experiences gap opening \citep[see the review by][]{LA2000}. In a one dimensional idealisation, the prograde disc is therefore often described as a {\it decretion} disc \citep{Pringle1991} in which there is a central torque and no gas flow onto the central binary, rather than a standard {\it accretion} disc in which there is no central torque and there is gas flow onto the central object \citep{PR1972,Pringle1981}. However, this idealisation of a decretion disc is often not realized in multi-dimensional studies. Although a tidally produced gap does form, gas flows through the gap in the form of streams, resulting in substantial accretion onto the binary \citep{AL1996,GK2002,MM2008,Shietal2012}. The accretion rates can be comparable to the rates expected from an accretion disc that surrounds a point mass whose mass is equal to the binary mass. The gap is formed by the resonant interaction of the binary with the disc \citep{AL1994}. 

In contrast to the prograde case, a retrograde circumbinary disc experiences no resonances and therefore flows towards the binary as a standard accretion disc until the disc orbits are perturbed significantly by close passages with the binary. The perturbed material is then captured into circumprimary or circumsecondary discs \citep{Nixonetal2011a}. For a circular orbit binary in a coplanar circumbinary Keplerian disc,  Lindblad resonances occur where 
\begin{equation}
\label{pp77}
\Omega^2(r) = m^2(\Omega(r)-\Omega_{\rm b})^2\,,
\end{equation}
where $\Omega_{\rm b}$ is the binary orbital frequency, $r$ is the distance from the binary center of mass in the binary orbit plane, $\Omega(r)$ is the disc orbital frequency, and $m>0$ is the integer wave mode number \citep[e.g.,][]{GT1979}. For a retrograde disc, the two frequencies have opposite signs and the right hand side of Equation (\ref{pp77}) is always larger than the left hand side. Therefore there can be no resonances in the disc.

In the case of eccentric orbit binaries, many more resonances are produced whose associated torques depend on the binary eccentricity \citep{GT1980}. Such resonances are capable of opening central gaps in prograde coplanar circumbinary discs \citep{AL1994}, while allowing gas flow through the gap in the form of time-modulated gas streams \citep{AL1996}. However, as we shall see below, when the binary is eccentric, it is possible to generate resonances between the binary and a counter-rotating disc. This effect is due to components of the eccentric binary potential that rotate in the opposite sense to the binary, and therefore progradely with respect to the disc. We describe this effect analytically and explore it with three dimensional hydrodynamic simulations. 

Retrograde circumbinary discs \citep{Nixonetal2011a,Nixon2012,RS2014} have received relatively little attention compared to prograde discs \citep[e.g.][]{AL1996,Escalaetal2005,Lodatoetal2009,Cuadraetal2009,Roedigetal2012,DOrazioetal2013}. The simulations of \cite{Nixonetal2012a} and \cite{Nixon2012} focussed on circular orbit binaries, with only a few simulations with $e \ne 0$, in which the eccentricity was always small. \cite{RS2014} performed simulations of retrograde discs with eccentricities as high as $e=0.9$, but due to the strongly self-gravitating discs employed, and the induction of disc tilt due to the dominant disc angular momentum \citep[cf.][]{Kingetal2005,Nixonetal2011b}, the subtle resonant effects sought here may have been missed. In this paper we focus on the case where the disc mass is negligible and so the binary dominates the system angular momentum, even in the highest eccentricity cases.

We present an analytic model and simulations of retrograde circumbinary discs, varying both the binary mass ratio and eccentricity. 
In Section \ref{mode_decomp} we define mode strengths that are an extension of the definition given by
\cite{Lubow1991b} to handle retrograde discs.   
 In Section \ref{sec:torque}, we evaluate by means of linear theory the torques
produced in a retrograde disc due to an eccentric orbit binary.
Section \ref{sec:sim} describes 
a set of simulations and their agreement with the analytic model,  Section \ref{sec:discussion} contains a discussion, and  Section \ref{sec:conclusions}
contains the conclusions.

\section{Mode Strengths}
\label{mode_decomp}

In order to gain insight into the response of a disc to tidal forcing, we decompose this response
into different modes that are characterized by azimuthal mode numbers $m$ and frequency mode
numbers $\ell$.
We consider a surface density distribution $\Sigma(r, \theta, t)$ and write
\begin{equation}
\Sigma(r, \theta, t) =   \sum \limits_{\ell=-\infty}^\infty \sum \limits_{m=0}^\infty Re[\Sigma_{\ell, m}(r) \exp{[i (m \theta - \ell  \tau)] \,]},
\label{sigsum}
\end{equation}
where $\Sigma_{\ell, m}(r)$ is a complex function and  $\tau = \Omega_{\rm b} t$ is the mean anomaly of the binary orbit that may be eccentric,
with mean motion $\Omega_{\rm b}>0$ (defined as $2 \pi/P_{\rm b}$ for binary orbital period $P_{\rm b}$). We invert Equation (\ref{sigsum}) to obtain
\begin{equation}
\Sigma_{\ell, m}(r)  = \frac{1}{2 \pi^2 (1 + \delta_{\ell, 0} \delta_{m,0})} \int_0^{2 \pi} \int_0^{2 \pi} \Sigma(r, \theta, t)
 \exp{[- i (m \theta - \ell  \tau)]} \, d\theta \, d\tau.
\label{siglm}
\end{equation}

Writing the integral in Equation (\ref{siglm}) in terms of real functions, we obtain
\begin{equation}
\Sigma_{\ell, m}(r) =  \Sigma_{\rm{cos, cos}, \ell, m}(r) +  \Sigma_{\rm{sin, sin}, \ell, m}(r) + 
 i \, (\Sigma_{\rm{sin, cos}, \ell, m}(r) -  \Sigma_{\rm{cos, sin}, \ell, m}(r))
\end{equation}
where
\begin{equation}
 \Sigma_{\rm{f, g}, \ell, m}(r) = \frac{1}{2 \pi^2 (1 + \delta_{\ell, 0} \delta_{m,0})} \int_0^{2 \pi} \int_0^{2 \pi} \Sigma(r, \theta, t) f(m \theta) g(\ell \tau)  \, d\theta \, d\tau,
\end{equation}
and $f$ and $g$ each can be the $\cos$ and $\sin$ functions. We define a radially integrated quantity $X$ as $\tilde{X}$ so that
\begin{equation}
\tilde{\Sigma}_{\ell, m} = \int_0^{\infty} \Sigma_{\ell, m}(r) \,r \, dr = \tilde{\Sigma}_{\rm{cos, cos}, \ell, m}+  \tilde{\Sigma}_{\rm{sin, sin}, \ell, m} + i \, (\tilde{\Sigma}_{\rm{sin, cos}, \ell, m}  -  \tilde{\Sigma}_{\rm{cos, sin}, \ell, m}),
 \label{Siglmreim}
\end{equation}
where
\begin{equation}
\tilde{\Sigma}_{\rm{f, g}, \ell, m} =  \int_0^{\infty} \Sigma_{\rm{f, g}, \ell, m}(r) \,  r \,dr \,.
\end{equation}
We then determine the square of the radially integrated mode amplitudes that are given by $|\Sigma_{\ell, m}|^2$ as
\begin{eqnarray}
\label{A}
|\tilde{\Sigma}_{\ell, m}|^2 &=&  (\tilde{\Sigma}_{\rm{cos, cos}, \ell, m} +  \tilde{\Sigma}_{\rm{sin, sin}, \ell, m})^2 + (\tilde{\Sigma}_{\rm{sin, cos}, \ell, m}  -  \tilde{\Sigma}_{\rm{cos, sin}, \ell, m})^2 \\
                    &=&  \tilde{\Sigma}_{\rm{cos, cos}, \ell, m}^2 +  \tilde{\Sigma}_{\rm{sin, sin}, \ell, m}^2+ \tilde{\Sigma}_{\rm{cos, sin}, \ell, m}^2 +  \tilde{\Sigma}_{\rm{sin, cos}, \ell, m}^2 + 2 \left( \tilde{\Sigma}_{\rm{cos, cos}, \ell, m} \tilde{\Sigma}_{\rm{sin, sin}, \ell, m} -  
      \tilde{\Sigma}_{\rm{sin, cos}, \ell, m} \tilde{\Sigma}_{\rm{cos, sin}, \ell, m} \right). \nonumber
\end{eqnarray}

We define the dimensionless mode strength $S_{\ell, m}$ such that $S_{0,0}$ is unity
\begin{equation}
S_{\ell, m} =  \frac{2 \pi \, |\tilde{\Sigma}_{\ell, m}|}{M_{\rm d}},
\label{S}
\end{equation}
where $M_{\rm d}$ is the disc mass. This definition of mode strength is similar to that given in \cite{Lubow1991b}. One difference is that  the 
definition in \cite{Lubow1991b} included only the squared terms in the second line of Equation (\ref{A}) and omitted the cross terms. To see the effect of the omission, consider the case of  density distribution
\begin{equation} 
\Sigma(r,\theta,t) = \delta(r-1) \cos{(m' \theta - \ell' \Omega_{\rm b} t )}.
\label{S1}
\end{equation}
In that case, we have that
\begin{equation}
|\tilde{\Sigma}_{\ell, m}|^2 =  \delta_{\ell,\ell'} \delta_{m,m'}.
\end{equation}
Using only the squared terms on the second line of Equation (\ref{A}), as in \cite{Lubow1991b}, we find that
\begin{equation}
\label{almnot}
|\tilde{\Sigma}_{\ell, m}|^2 - 2 \left( \tilde{\Sigma}_{\rm{cos, cos}, \ell, m} \tilde{\Sigma}_{\rm{sin, sin}, \ell, m} -  
      \tilde{\Sigma}_{\rm{sin, cos}, \ell, m} \tilde{\Sigma}_{\rm{cos, sin}, \ell, m} \right) = \frac{1}{2}\delta_{|\ell| , |\ell'|} \delta_{m,m'}.
\end{equation}
Equation (\ref{almnot}) shows that the previous
definition of mode strengths does not distinguish between prograde and retrograde modes in which the sign of $\ell$ differs. It was adequate for the analysis of the superhump instability cycle that only involved prograde ($\ell>0$) modes, but is not adequate for the current study. Instead we must use the definition of $S_{\ell, m}$ given by Equation (\ref{S}). In Appendix \ref{sec:mssph}, we describe how mode strengths are computed in SPH simulations.

\section{Resonant Torques}
\label{sec:torque}

\subsection{Torque equation}
We determine the resonant torque on a gas disc due to an eccentric binary that is exerted on a counter-rotating disc.  
The disc mass is assumed to be very small compared to the binary mass.
We take the binary to be on a prograde orbit with mean motion $\Omega_{\rm b}>0$, while the disc orbit is prograde $\Omega(r)>0$ or retrograde $\Omega(r) <0$ with respect to the binary. We adopt a cylindrical coordinate system whose origin is at the binary center of mass and so 
the $z=0$ plane coincides with the binary orbit plane. Following \cite{GT1979}, we decompose the potential as
\begin{equation}
\Phi(r, \theta, t) = \sum_{\ell, m} \Phi_{\ell, m}(r) \cos{(m \theta - \ell \Omega_{\rm b} t)}\,,
\label{Phi}
\end{equation}
where $m \ge 0$ and $\ell$ ranges over all integers (negative, zero, and positive). The binary potential has contributions from the primary star of mass $M_1$ and the secondary star of mass $M_2$. We determine the potential components $\Phi_{\ell, m}$ for $\ell \ne 0$ and $m \ge 0$, by  
\begin{equation}
 \Phi_{\ell, m}(r) = \Phi^{(1)}_{\ell, m}(r)+ \Phi^{(2)}_{\ell, m}(r)
\end{equation}
for $m>0$ where the superscripts 1 and 2 denote the primary and secondary objects. Following \cite{MA2008}, we invert Equation~(\ref{Phi}) by using the eccentric anomaly of the binary orbit $\zeta$ as a variable of integration in place of the mean anomaly, $\Omega_{\rm b} t$. We then have for $i=1,2$
\begin{equation}
\Phi^{(i)}_{\ell, m}(r) = -\frac{G M_i}{2 \pi^2 a} \left(1- \frac{\delta_{m,0}}{2} \right) \int_0^{2 \pi}d \theta  \int_0^{2 \pi} d \zeta\,\, \frac{\left(1- e \cos{\zeta}\right)\cos{( m \theta -\ell (\zeta - e \sin{\zeta}) )}}{\sqrt{\beta^2 + x_i^2 (1-e  \cos{\zeta})^2- 2 \beta x_i g(\theta,\zeta)}},
\end{equation}
where
\begin{eqnarray}
g(\theta,\zeta) = (\cos{\zeta} -e) \cos{(\theta)}+\sqrt{1-e^2} \sin{(\zeta)} \sin{(\theta)},
\end{eqnarray}
$x_1 = - M_2/(M_1+M_2)$ and $x_2=M_1/(M_1+M_2)$, $e$ is the binary eccentricity,  and $\beta=r/a$, for a binary with semi-major axis $a$. The advantage of this method is that we can determine $\Phi^{(i)}_{\ell, m}$ numerically for arbitrary binary eccentricity. We have verified analytically in expansions for small $e$ that this method recovers potentials $\Phi_{1,1}$, $\Phi_{1,2}$, and $\Phi_{1,3}$ given by Equations (18), (21), and (23) respectively of \cite{AL1994}.

We determine the resonant torque on the circumbinary disc due to outer Lindblad resonances that occur where
\begin{equation}
m \Omega(r) - \ell \Omega_{\rm b} = -\kappa(r),
\label{Omr}
\end{equation}
where $\kappa$ is the epicyclic frequency. Note that we do not assume the disc is Keplerian. We take into account the non-Keplerian effects of the binary gravitational potential by determining $\Omega(r)$ and $\kappa(r)$ as
\begin{equation}
\Omega^2(r) = \frac{1}{r}\frac{d \Phi_{0, 0}}{d r}
\label{Om}
\end{equation}
and 
\begin{equation}
\kappa^2(r) = \frac{3}{r}\frac{d \Phi_{0, 0}}{d r} + \frac{d^2 \Phi_{0, 0}}{d^2 r},
\label{kappa}
\end{equation}
where we take $\Omega$ and $\kappa$ to be negative for a counter-rotating disc.

In the case of a Keplerian disc, we have that $\Omega=\kappa$ and so
\begin{equation}
 \Omega(r) = \frac{\ell \Omega_{\rm b}}{m+1}
\label{Omrk}
\end{equation}
at an outer Lindblad resonance. This estimate treats the binary as a point mass which is valid at large distances from binary. However, as noted above, we do not make this approximation in the torque calculations that we carry out.

The magnitude of the resonant torque on the disc \citep{GT1979} is given by 
\begin{equation}
T_{\ell,m} = \frac{ m \pi^2 \Sigma \Psi_{\ell,m}^2}{|\cal D|} ,
\label{Tlm}
\end{equation}
where 
\begin{equation}
\Psi_{\ell,m} = r \frac{d \Phi_{\ell, m} }{dr}+ \frac{2m \Omega}{m \Omega(r)-\ell \Omega_{\rm b}}  \Phi_{\ell, m}
\end{equation}
and 
\begin{equation}
{\cal D} = r  \frac{d(\kappa^2 - (m \Omega- \ell \Omega_{\rm b})^2)}{dr}.
\label{calD}
\end{equation}
In Equations (\ref{Tlm}) - (\ref{calD}), all radially dependent quantities are evaluated at the resonance radius given by Equation (\ref{Omr}). We find that the outer Lindblad resonance torques we consider lead to pushing material outward for both prograde and retrograde discs.

\subsection{Results}

We describe here the results of the application of the torque model to the cases that we have analysed in the simulations discussed later in Section \ref{sec:sim} that have $(\ell, m)$ values of (1,1), (1,2), (-1,0), (-1,1), and (-1,2). From Equation (\ref{Omrk}),  it is then clear that torques with negative $\ell$ values are possible in counter-rotating discs for eccentric binaries, since $\Omega$ is negative at resonance. However, since $m \ge 0$, it then follows that such resonances involve $\ell < 0 \le m$. \cite{GT1980} showed that $T_{l,m} \propto e^{2 |\ell-m|}$ for $e \ll 1$. That is, resonant torques are possible in a counter-rotating disc for eccentric binaries ($e > 0$) provided that $\ell <0< m$, but they are generally weaker for larger differences between $m$ and $\ell$. We then expect  that the  strongest Lindblad torques that are produced in a disc that counter-rotates with respect to the sense of orbital rotation of a binary to have $\ell=-1$.  For this reason, we have concentrated on resonances with $\ell=-1$.
 
For the cases of disc modes $(\ell, m)$ of (1,1) and (1,2), we see from the approximate Equation (\ref{Omrk}) that resonances occur only for $\Omega(r) >0$. Since the simulated discs are retrograde $\Omega <0$, we expect that there are no resonant torques for these simulated cases. Therefore, we predict that $S_{1,1}$ and $S_{1,2}$ are both zero. These cases serve as as useful check on the both the theory and the numerical accuracy of the mode strength determinations from simulations.

For the case of disc mode $(-1, 1)$, Equation (\ref{Omrk}) estimates that the resonance is located at $r=1.58 a$. This resonance is sufficiently close to the binary that nonKeplerian effects can be important. In addition, the binary may cause the disc to be dynamically unstable. Consider first the case of a circular binary. If we approximate the disc as consisting of simple periodic orbits in the corotating frame of the binary, we can apply the previously known results on ballistic particle orbits in the restricted three-body problem to study the existence and stability of disc streamlines near the resonance. A similar approach for the prograde case with a circular binary was taken by \cite{Paczynski1977}. Circumbinary, retrograde, simple periodic orbits are known to be quite stable, even if they come close to the binary, although the orbits become increasingly noncircular close to the binary \citep{Szebehely1967}. We have found that there are only mild departures from circular orbits near this resonance location in the case of a circular binary.

\begin{figure}
  \center{\includegraphics[width=0.7\textwidth]{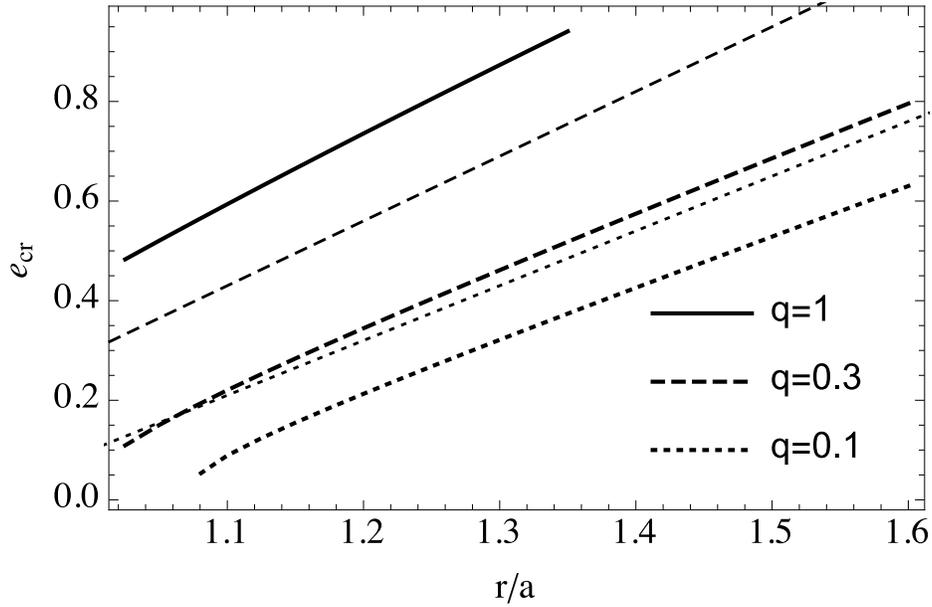}}
  \caption{The heavy lines plot critical eccentricity values for which the epicyclic frequency $\kappa$ vanishes as a function of radius. The light lines plot the critical eccentricity values for which the apocentre of the secondary object reaches a radius on the horizontal axis. The  solid, dashed, and dotted lines are  for binary mass ratios of $q=1$, $q=0.3$, and $q=0.1$, respectively. No light solid line is plotted because a member of an equal mass binary cannot reach the plotted radii ($r/a > 1$) for any value of eccentricity.}
  \label{ecr}
\end{figure}

In the case that the binary is eccentric, the situation is somewhat different. Due to the non-Keplerian effects of the binary, the epicyclic frequency $\kappa$ is a function of binary eccentricity. At locations close to the binary, $\kappa$ can vanish for sufficiently high binary eccentricity. In Figure \ref{ecr}, we plot as heavy lines the critical values of binary eccentricity $e_{\rm cr}$ for which $\kappa$ vanishes as a function of distance from the binary center of mass. The plot shows that $e_{\rm cr}$ is smaller closer to the binary, as one would expect. For eccentricities above the line, $\kappa^2$ is negative which means that the epicyclic motions are unstable and resonance is not possible. The equal mass binary case is more stable than the unequal mass cases, possibly because the binary members are located farther inward from a given radius in the circumbinary disc. 
 
Also in Figure \ref{ecr}, we plot in light lines the critical eccentricity for which the apocentre of the secondary object reaches a given radius. Above these lines, the secondary member of the binary would cross the given radius. Consequently, disc-like orbits should not exist at such radii. For an equal mass binary, the apocentre reaches a radius $r/a =1$ when $e=1$. The binary cannot extend beyond $r/a >1$ for any value of $e \le1$. Consequently, we do not plot a light solid line in the figure. The light lines lie above the corresponding heavy lines. The plot then shows that for such critical binary eccentricities with $1 \ge q \ge 0.1$, the disc orbit is already unstable, since $\kappa^2 <0$.
 
The angular speed  $\Omega(r)$, as well as $\kappa(r)$, is affected by the binary eccentricity. As a consequence, the resonance location given by Equation (\ref{Omr}) changes with binary eccentricity. For the (-1,1) resonance, we find that the effect of binary eccentricity is to pull the resonance inward of the Keplerian location of $1.58 a$. As a result, the critical value for the eccentricity at this resonance is even smaller than would be predicted by Figure \ref{ecr} for the Keplerian location. We find that the critical value of eccentricity for a binary of mass ratio $q=0.3$ is $e_{\rm cr} \simeq 0.55$. Above the value, we expect $T_{-1,1}$ and $S_{-1,1}$ to vanish for $q=0.3$. In addition, from symmetry considerations, it it easy to see that $\Phi_{-1,1}=0$ for $q=1$. Therefore,  $T_{-1,1}$ and $S_{-1,1}$ should vanish in the case of an equal mass binary.

Similar stability considerations apply to the resonance for (-1,0) that lies even closer to the binary. In this case, the torque is always zero because $m=0$ for this mode. However, an energy flux can be generated at such a resonance and $S_{-1,0}$ can in principle be nonzero. Equation (\ref{Omrk}) suggests that the resonance occurs where $\Omega=-\Omega_{\rm b}$, if the disc were Keplerian there. Due to the strong nonKeplerian effects, we expect that the resonance condition is even more difficult to satisfy in this case than in the (-1,1) case. Consequently, we expect $S_{-1,0}$ to also be small.

\begin{figure}
  \center{\includegraphics[width=0.7\textwidth]{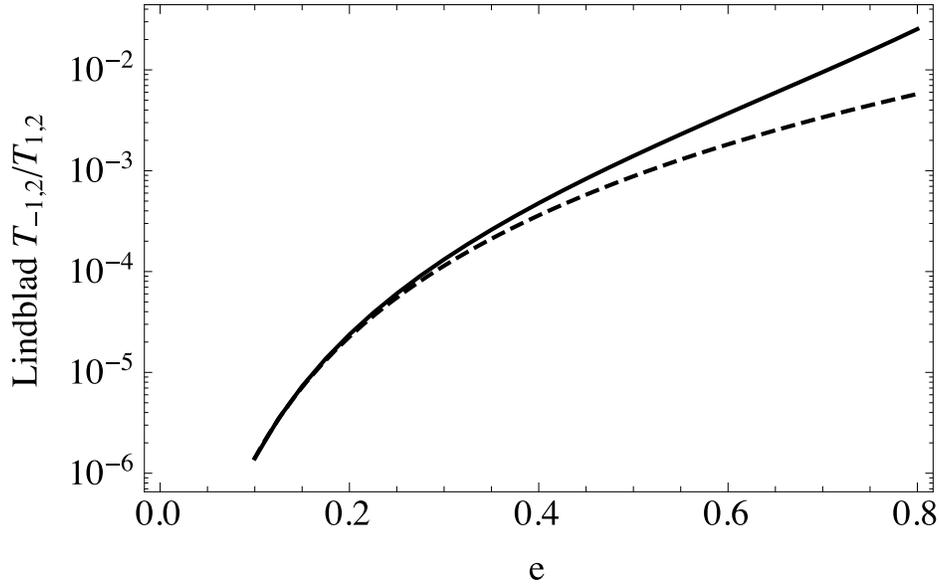}}
  \caption{Torque ratio of the resonant torques $T_{-1,2}/T_{1,2}$ at the $\mp$1:3 eccentric outer Lindblad resonance for an equal mass binary, assuming the same disc density at the resonance in the two cases of a retrograde and prograde disc, respectively. The solid line is the result of numerical evaluation of Equation \ref{Tlm}. The dashed line follows $\propto e^4$ that is expected to be valid for $e \ll 1$. }
  \label{TR}
\end{figure}

\begin{figure}
  \center{\includegraphics[width=0.7\textwidth]{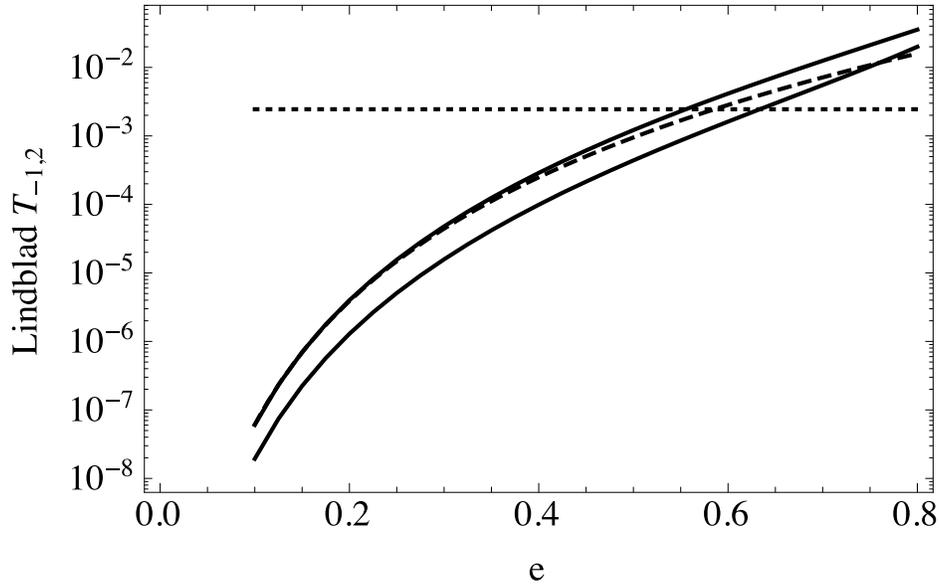}}
  \caption{Upper solid line plots torque $T_{-1,2}$ due to eccentric outer -1:3 retrograde Lindblad resonance  in units of $\Sigma \Omega_{\rm b}^2 r^4$ at the resonance as a function of binary eccentricity $e$ for an equal mass binary. The lower solid line is for a binary mass ratio of 0.1.  The dashed line follows $\propto e^6$ that is expected to be valid for $e \ll 1$. The horizontal dotted line plots the viscous torque at the resonance for the disc parameters used in the simulations.}
  \label{Tm12}
\end{figure}

Torque $T_{-1,2}$ occurs near $\Omega(r)=-\Omega_{\rm b}/3$ or $r \simeq 2.08 a$ that is far enough from the binary to permit resonances to operate. 
At this resonance, the disc is in nearly Keplerian motion. In this case,  for small $e$, we expect  that $T_{-1,2} \propto e^{2|l-m|} \propto e^6$ and the torque is consequently very sensitive to the binary eccentricity. 
In a prograde disc,  torque $T_{1,2}$ occurs at the same radial location. The prograde counterpart to torque $T_{-1,2}$  in a retrograde disc 
is then $T_{1,2}$. This  torque $T_{1,2}$ is frequently found to truncate the inner regions of  prograde discs that orbit around binaries with modest eccentricity $e \sim 0.1$ \citep{AL1994}. In Figure \ref{TR}, we plot the ratio  $T_{-1,2}/T_{1,2}$ as a function of eccentricity in the two cases of a retrograde ($T_{-1,2}$) and prograde ($T_{1,2}$) disc. Notice that the retrograde torque is quite weak compared to the prograde torque and there is a strong dependence on eccentricity that is expected to be $\propto e^4$ for $e \ll1$, as seen in the figure. In Figure \ref{Tm12}, we plot the  torque $T_{-1,2}$ and the viscous torque at the resonance for the disc parameters adopted in the simulations of Section \ref{sec:sim}. The dashed line shows that $T_{-1,2}$ follows the expected dependence on eccentricity $e^6$ for small $e$. For fairly high values for $e \sim 0.5$, the resonant torque can overpower the viscous torque and truncate the disc. 
 
In summary, of the disc mode strengths we consider, we expect $S_{1,1}$, $S_{1,2}$ and $S_{-1,0}$ to be small compared to $S_{-1,1}$ and $S_{-1,2}$. Whereas $S_{-1,1}$ can be significant for intermediate eccentricities that are less than about 0.55, it vanishes for equal mass binaries. $S_{-1,2}$ should dominate among these cases at higher eccentricities $e > 0.6$, especially for $q=1$.

\section{Simulations}
\label{sec:sim}
We present numerical simulations using the smoothed particle hydrodynamics \citep[SPH; e.g.][]{Price2012a} code {\sc phantom} \citep[e.g.][]{PF2010,LP2010}. This code has already been applied to many types of accretion in binary systems \citep{Nixon2012,Rosottietal2012,Nixonetal2013,Facchinietal2013,Martinetal2014a,Martinetal2014b}. The simulations include the effects of disc pressure and viscosity, while effects of disc self-gravity are ignored. The binary is modelled as two Newtonian point masses which affect the gas orbits through gravity and accrete any particles which come close enough. The back reaction on the binary orbit is also included, but the disc mass is too small to result in any significant changes on the simulation timescale.

\subsection{Setup}

We set up a binary with mass ratio $q = M_2/M_1$ and total mass $M = M_1+M_2 = 1$, semimajor axis $a=1$, and eccentricity $e$ initially at apocentre of the orbit. The binary is taken to rotate with positive angular speed and so rotate with positive mean motion  $\Omega_{\rm b}$ (defined as $2 \pi/P_{\rm b}$ for binary orbital period $P_{\rm b}$), while the disc rotates in the opposite sense with $\Omega <0$. The binary sink particles have accretion radii of $0.2a$, inside which gas is removed from the simulation and its mass and momentum added to the sink particle. We initialise the disc with 8 million particles in Keplerian orbit about the binary centre of mass. The disc initially extends from an inner radius of $2a$ to an outer radius of $8a$ with a surface density distribution of $\Sigma(r) = \Sigma_0 (r/r_0)^{-3/2}$, where the value of $\Sigma_0$ is set to provide a disc mass of $M_{\rm d} = 10^{-3}M$. Throughout the simulations the binary accretes $\lesssim 10$\% of the disc. The simulations adopt the locally isothermal assumption in which the disc sound speed is $c_{\rm s}(R) = c_{{\rm s,}0}(R/R_0)^{-3/4}$ for spherical radius $R$ and $c_{\rm s,0}$ corresponds to aspect ratio $H/R = 0.05~(0.035)$ at the disc inner (outer) edge at all times. Since the discs are thin, the difference between between the cylindrical radius $r$ and spherical radius $R$ at any point in the main body of the disc is small. We employ an explicit accretion disc viscosity which corresponds to an approximately constant \cite{SS1973} $\alpha=0.05$ throughout the initial disc \citep[Section 3.2.3 of][]{LP2010}. The viscous stresses include a nonlinear term with a coefficient $\beta_{\rm AV}=2$ (AV stands for artificial viscosity) that suppresses inter-particle penetration, as is standard in SPH codes. For this configuration, the shell averaged smoothing length per disc scale--height \citep[see][]{LP2010} is $\left<h\right>/H \approx 0.16$ for $8$ million particles. The simulations reported below span $q \approx 0.1, 0.3,$ and $1.0$ (more precisely $9\colon\!1$, $3\colon\!1$ and $1\colon\!1$) and $e = 0.0,~0.2,~0.4,~0.6$, and $0.8$. 

\subsection{Results}
\begin{figure*}
  \begin{center}
    \includegraphics[angle=0,width=\textwidth]{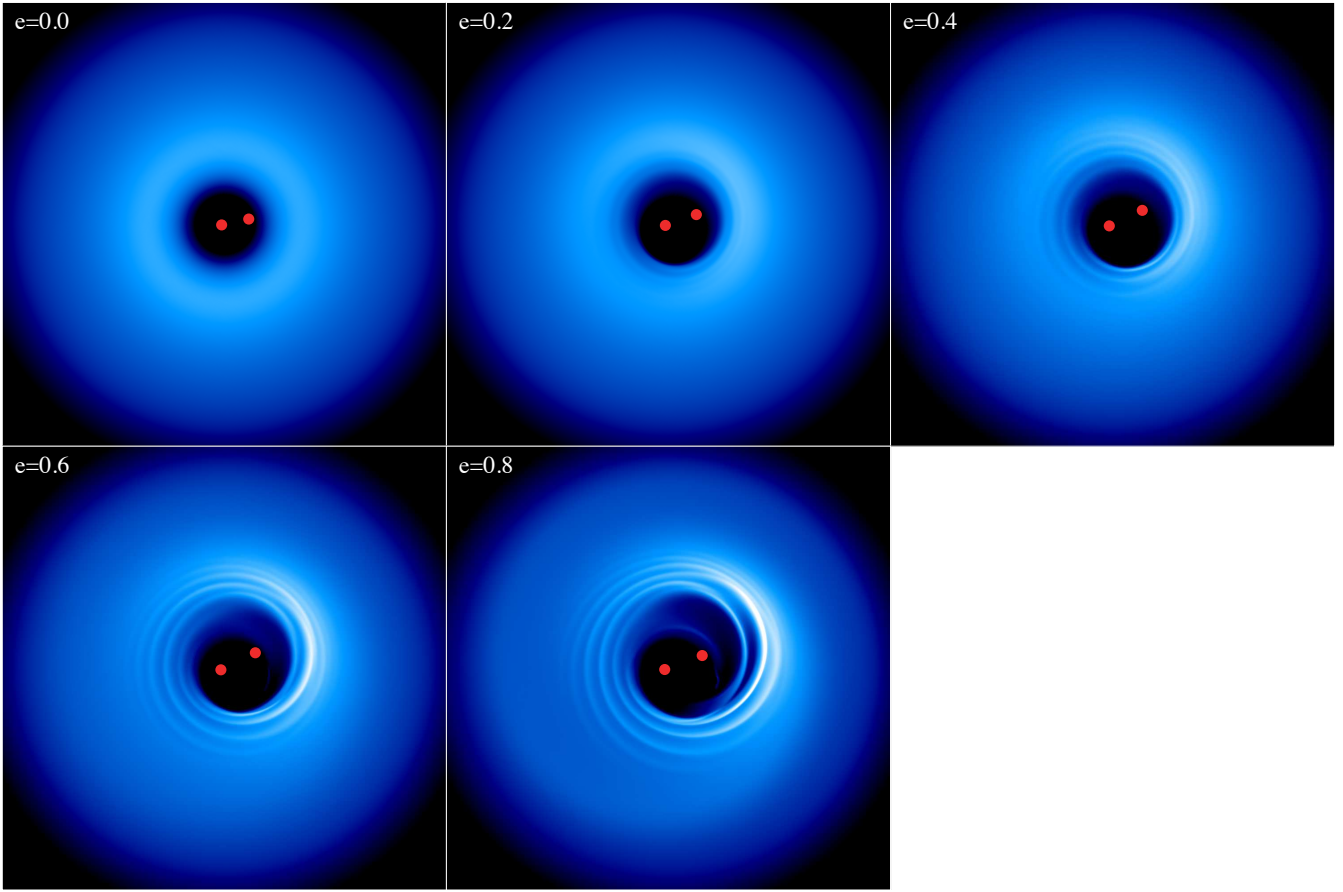}
    \caption{Surface density rendering of each simulation with $q = 0.1$ at $t=628$ (after 100 binary orbits). The binary rotates in a counter-clockwise sense and the disc rotates in a clockwise sense. The binary eccentricity is shown on each panel. The colour scheme, from lowest surface density (black) to highest surface density (white) covers approximately 2 orders of magnitude. The binary is represented by the two red filled--circles, with the circle size denoting the accretion radius inside which particles are removed from the simulation.}
    \label{fig1}
  \end{center}
\end{figure*}

\begin{figure*}
  \begin{center}
    \includegraphics[angle=0,width=\textwidth]{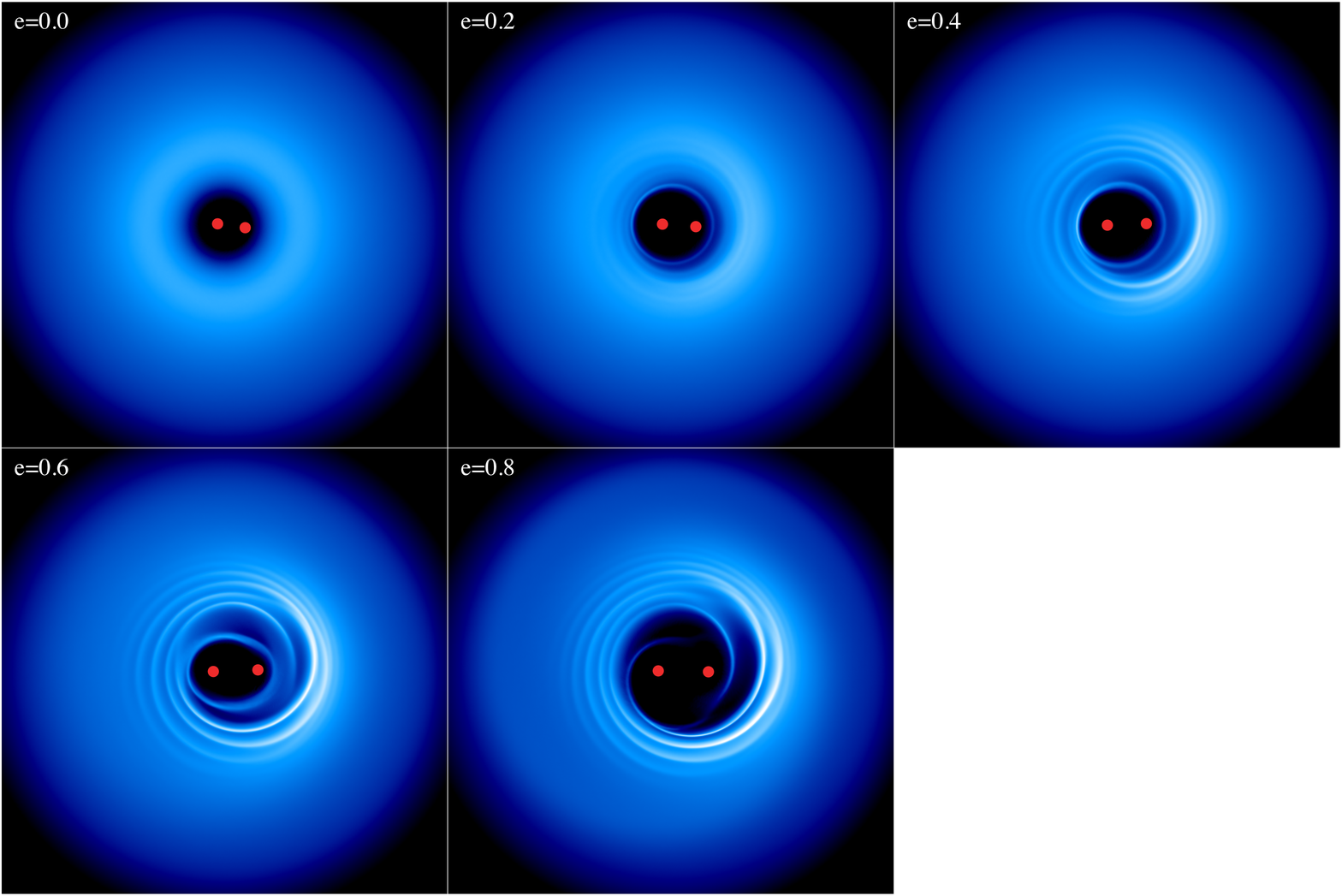}
    \caption{Same as Figure~\ref{fig1}, but for $q = 0.3$.}
    \label{fig2}
  \end{center}
\end{figure*}

\begin{figure*}
  \begin{center}
    \includegraphics[angle=0,width=\textwidth]{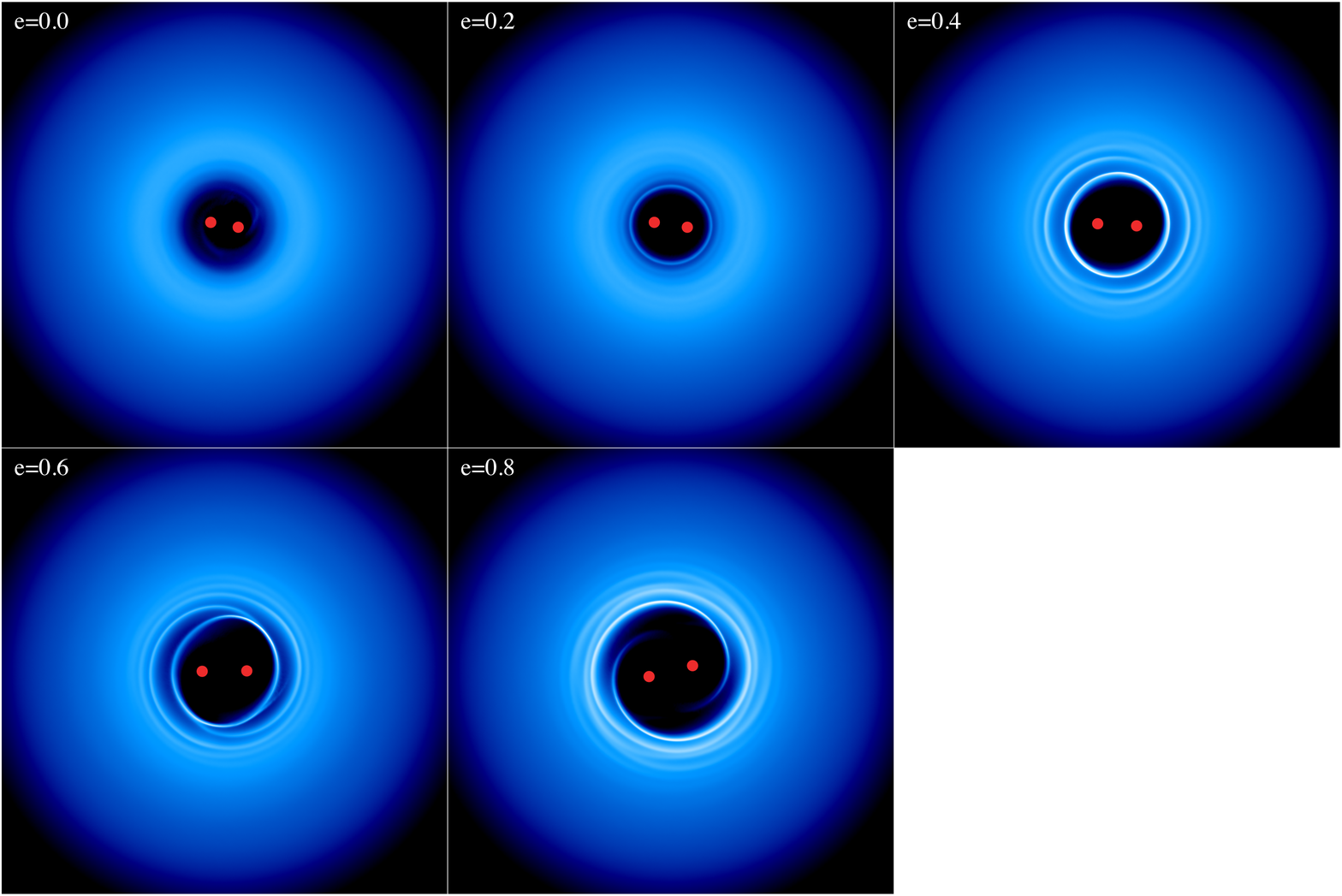}
    \caption{Same as Figure~\ref{fig1}, but for $q = 1.0$.}
    \label{fig3}
  \end{center}
\end{figure*}

In Figures~\ref{fig1}, \ref{fig2}, and \ref{fig3} we show the surface  density renderings of the simulations at their completion after 100 binary orbits. For each mass ratio the general results are the same: for a circular binary there are no disc structures that signify resonances between the disc and the binary, when the binary is eccentric the disc structures appear and generally grow stronger with increasing eccentricity. The unequal mass ratio simulations, $q = 0.1, 0.3$, show significantly asymmetric structures, whereas the equal mass ratio simulations (which begin with symmetric initial conditions) remain bisymmetric throughout the duration of the simulation. 

The form of the disc disturbances seen in the figures is of spiral waves that have $\chi=r {\rm d} \theta_{\rm s}/ {\rm d}r >0$, where $\theta_{\rm s}(r)$ is the azimuthal angle of the spiral. This sign of $\chi$  is expected from the theory of disc resonances in Section \ref{sec:torque} because each resonance is due to a retrograde bar that excites ``trailing'' waves in the retrograde disc. We note that if the disc were responding to a prograde disturbance, the sign of the $\chi$ would be negative. Only if the retrograde disc is responding to a disturbance that rotates in a retrograde sense at a faster speed than the gas can $\chi$ be positive, as we find. That condition occurs at an outer Lindblad resonance due to a retrograde bar. In addition, the gas streams in the gap are also expected to have $\chi>0$ because the streams approximately conserve angular momentum as they flow towards the binary. The gas then rotates faster in the retrograde sense, resulting in $\chi>0$.

\begin{figure*}
  \begin{center}
    \includegraphics[angle=0,width=0.5\textwidth]{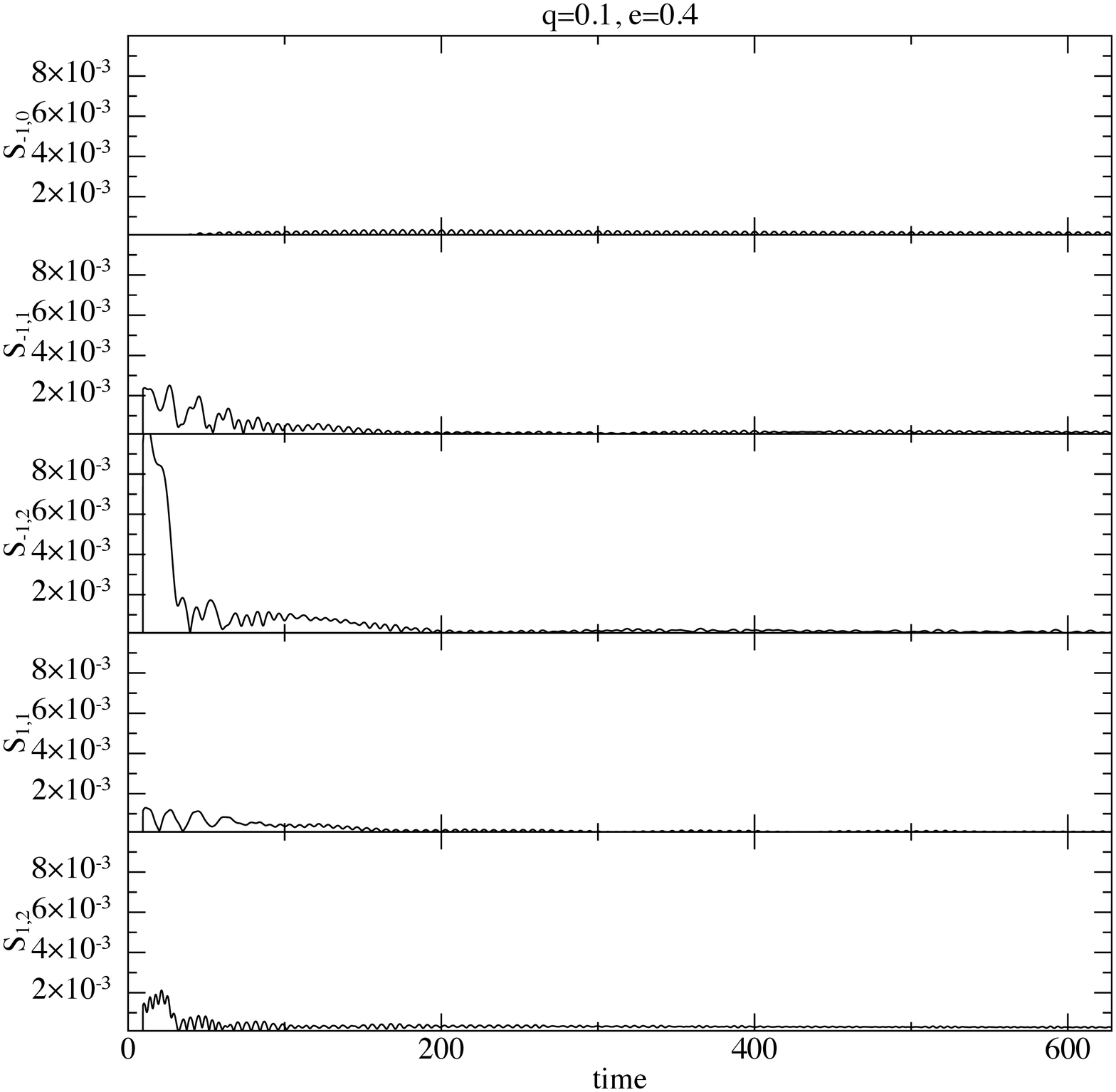}
    \caption{The different mode strengths as a function of time for the $q=0.1$, $e=0.4$ simulation. The time axis is in units where $2\pi$ is one binary orbital period.}
    \label{fullmodes1}
  \end{center}
\end{figure*}

\begin{figure*}
  \begin{center}
    \includegraphics[angle=0,width=0.5\textwidth]{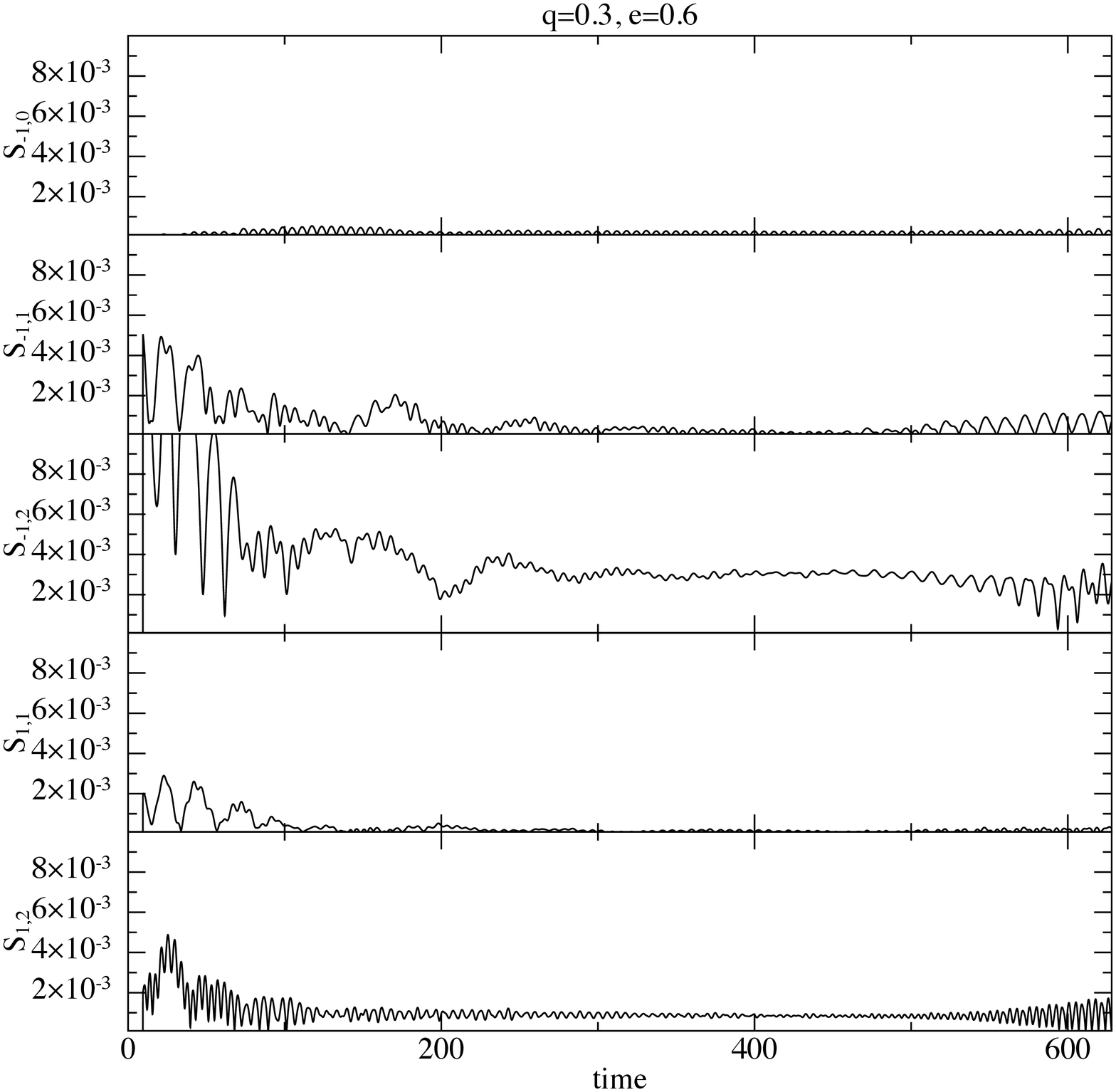}
    \caption{The different mode strengths as a function of time for the $q=0.3$, $e=0.6$ simulation. The time axis is in units where $2\pi$ is one binary orbital period.}
    \label{fullmodes2}
  \end{center}
\end{figure*}

\begin{figure*}
  \begin{center}
    \includegraphics[angle=0,width=0.5\textwidth]{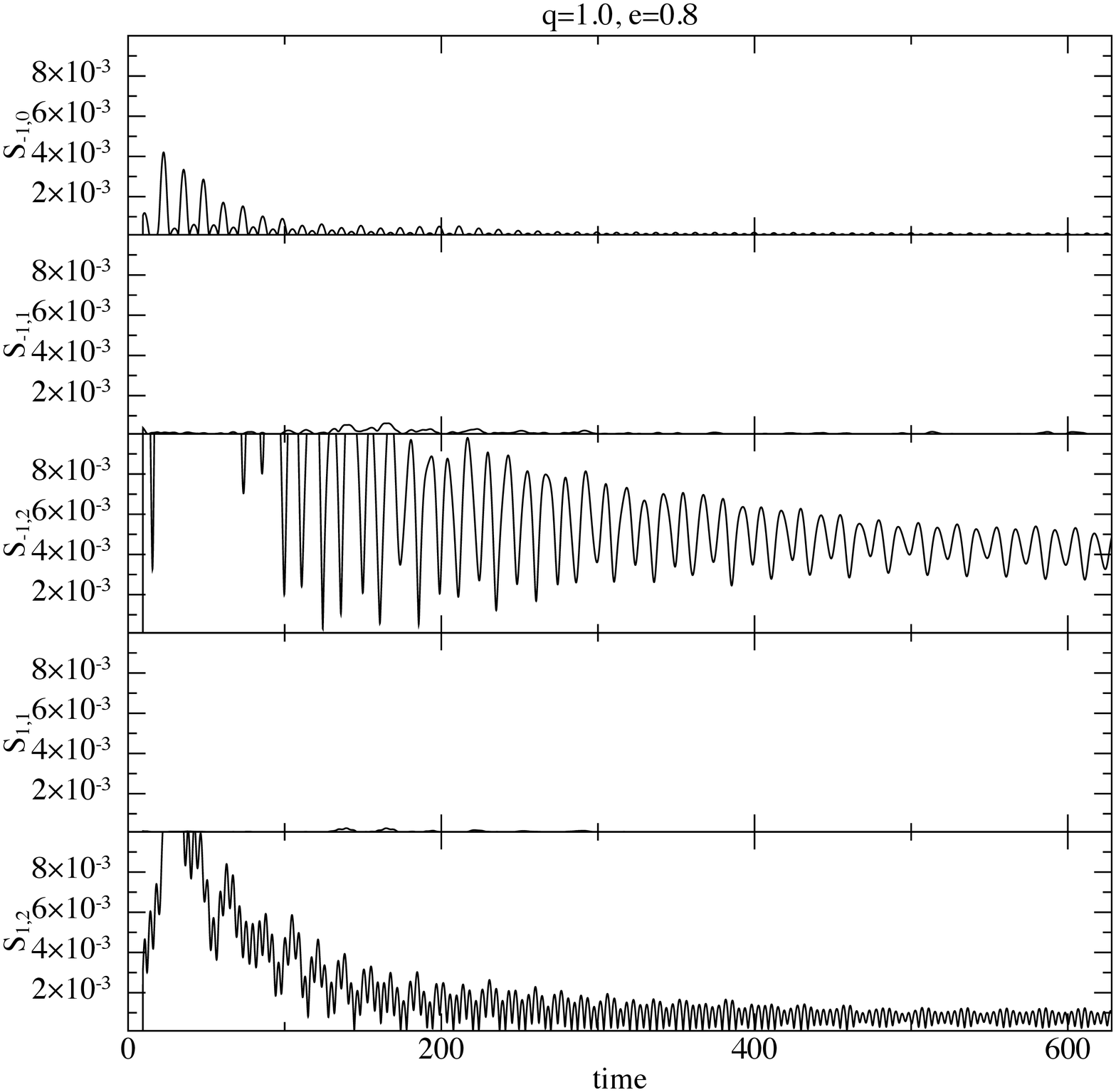}
    \caption{The different mode strengths as a function of time for the $q=1.0$, $e=0.8$ simulation. The time axis is in units where $2\pi$ is one binary orbital period.}
    \label{fullmodes3}
  \end{center}
\end{figure*}

The circular equal mass ratio simulation displays weak streams feeding the sink particles from the inner edge of the disc. Such features may be present in the unequal mass ratio cases, but here the secondary sink is closer to the disc inner edge and accretes the gas which might form them. 
In Figures~\ref{fullmodes1}, \ref{fullmodes2}, and \ref{fullmodes3} we show the mode strengths against time for a selection of the simulations. As the disc is not set up in equilibrium with the binary orbit, there is some initial settling of the disc which appears in the mode strengths. Figure~\ref{fullmodes1} shows the five calculated mode strengths for the $q=0.1$, $e=0.4$ simulation. Figure~\ref{fullmodes2} shows the five calculated mode strengths for the $q=0.3$, $e=0.6$ simulation. Figure~\ref{fullmodes3} shows the five calculated mode strengths for the $q=1.0$, $e=0.8$ simulation. These figures illustrate the general trends found in this work: (1) smaller mass ratios create weaker resonances, (2) higher eccentricities create stronger resonances, and (3) for strong enough resonances the $(-1,2)$ mode can prevent gas from reaching the $(-1,1)$ resonance on circular orbits, depleting its strength. In addition, the  $(-1,1)$  resonance is likely unstable at higher eccentricties, as discussed in Section \ref{sec:torque}.

The extent of the gap and the (-1, 2) mode strengths are fairly stable at the end of the simulations. However, since the simulations are limited to 100 binary orbits, we cannot be certain of the behaviour over the longer term evolution of the disc. The $(l,m) = (-1,2)$ mode strength is the largest and displays the clearest trend of increasing strength with increasing eccentricity, as expected by the analytic model in Section \ref{sec:torque}. The $(1,1)$ mode  strength is small for all parameters, while the $(1,2)$ mode  strength displays some fluctuations with increasing eccentricity. The (-1,1) mode strength nearly vanishes for an equal mass binary and is weaker than the  $(-1,2)$  mode strength. The behavior of the mode strengths is then in good agreement with the expectations of the analytic model discussed in Section \ref{sec:torque}.

\subsection{Binary evolution and accretion}
\label{sec:binev}
\cite{Nixonetal2011a} developed an analytic model for the interaction of a binary with a circumbinary retrograde disc. The secondary object is taken to be of small mass compared to the primary. The model assumed that the secondary object (or its surrounding disc) experiences an impact and accretion from the circumbinary gas that is locally counter-rotating at circular Keplerian speeds. In the model, the binary loses angular momentum by gravitational interaction with the retrograde gas. Its semi-major axis decreases, while it eccentricity increases, provided that its eccentricity is greater than a threshold value $\sim H/r$. 

In the case that the binary opens a gap in a retrograde circumbinary disc, the situation is somewhat different. The binary exerts a tidal torque at a resonance, such as the $(-1,2)$ resonance, due to a retrograde component of its potential that is prograde with the rotation of the disc, as discussed in Section \ref{sec:torque}.  The resonance has the effect of pushing gas outward, away from binary, which adds negative angular momentum to the $\Omega <0$ disc and therefore adds positive angular momentum to the binary.  The resonant torque on the disc is then negative and the torque on the binary is positive. In addition, the gas that impacts the secondary flows in the form of gas streams that originate at the disc inner edge. These streams are retrograde to the binary and should reduce the angular momentum on the binary. However, they are not in circular motion.  Therefore, it is unclear how well the model of \cite{Nixonetal2011a} applies to the case of a highly eccentric binary.

\begin{figure*}
  \begin{center}
    \includegraphics[angle=0,width=0.84\textwidth]{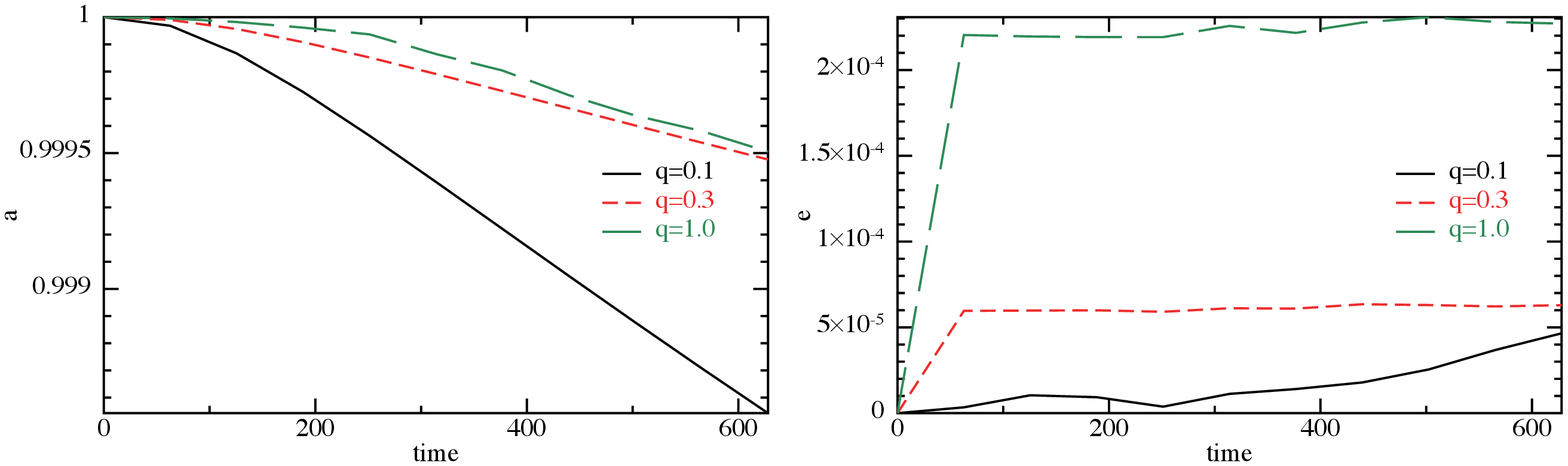}
    \includegraphics[angle=0,width=0.84\textwidth]{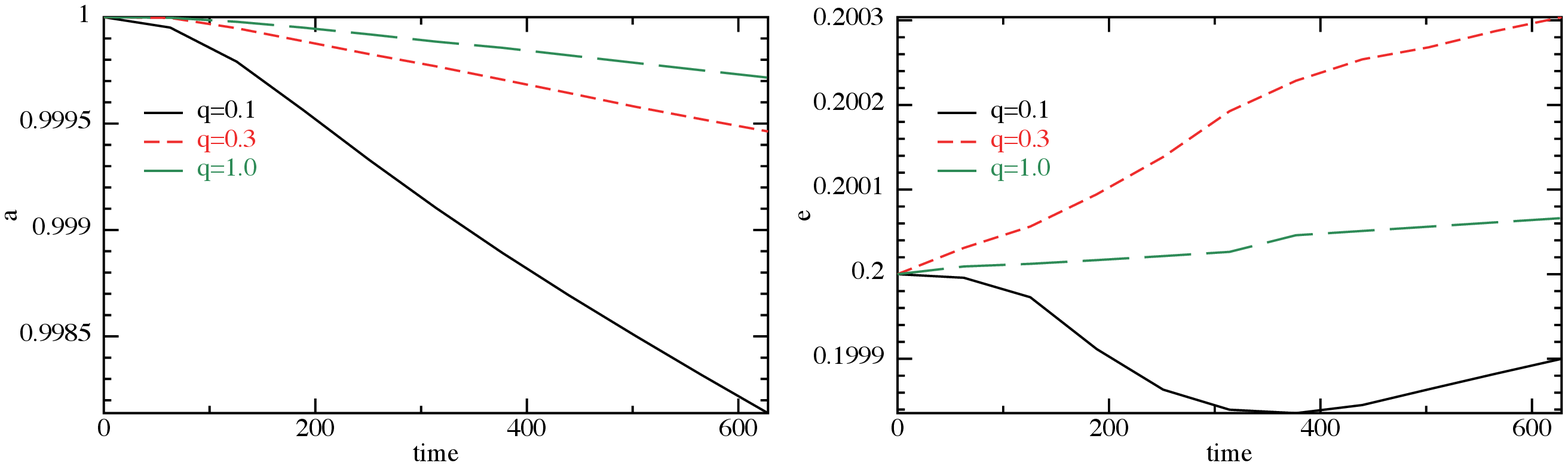}
    \includegraphics[angle=0,width=0.84\textwidth]{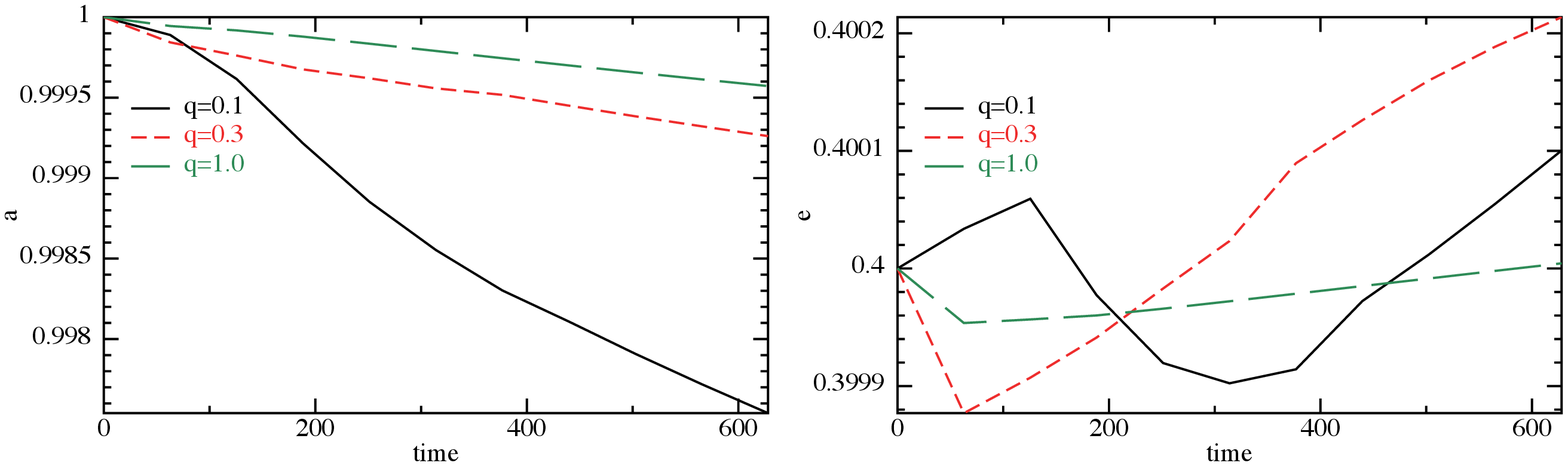}
    \includegraphics[angle=0,width=0.84\textwidth]{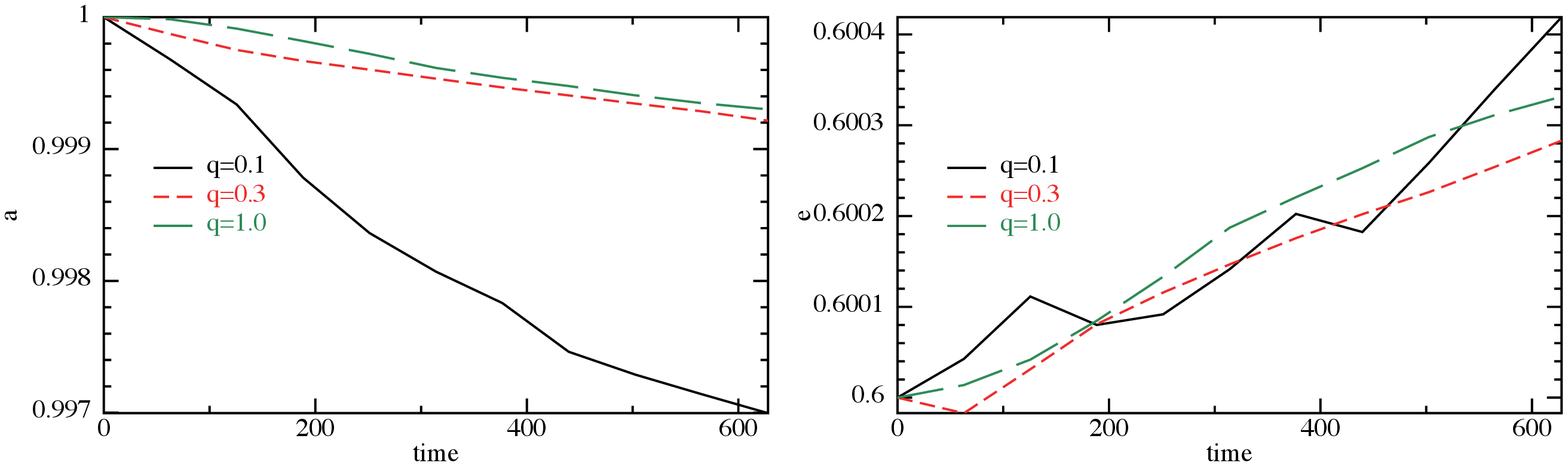}
    \includegraphics[angle=0,width=0.84\textwidth]{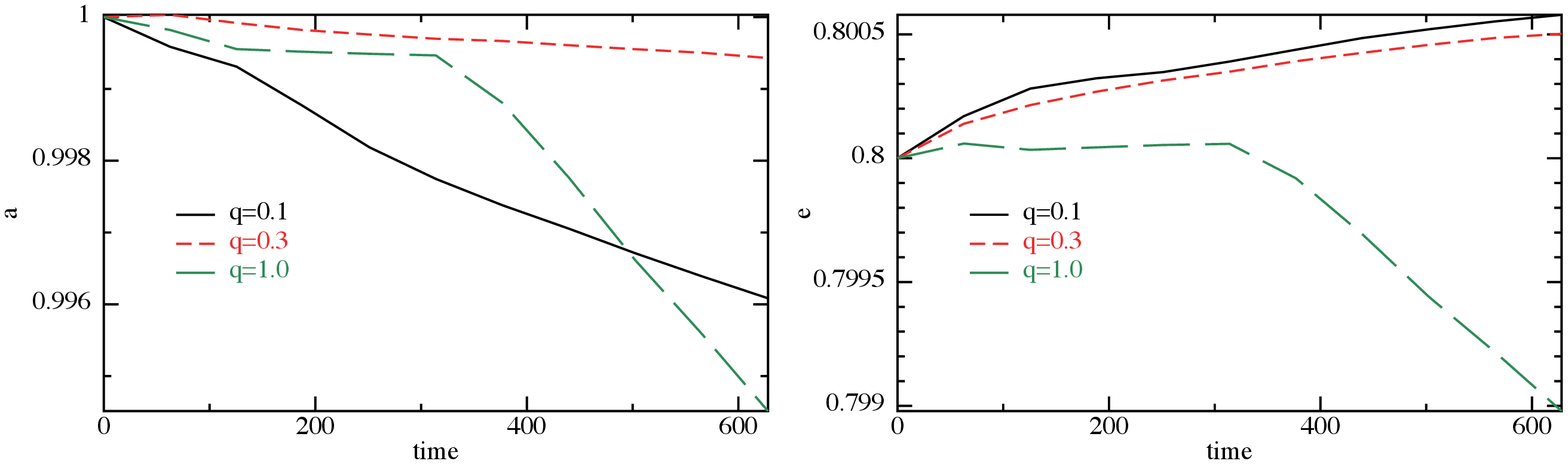}
    \caption{Evolution of the binary semi--major axis, $a$, and eccentricity, $e$. The time is in units where $2\pi$ is one binary orbital period. The semi--major axis is shown on the left hand panel and the eccentricity is shown in the right hand panel. From top to bottom the rows show the different initial eccentricities from $e = 0.0 - 0.8$. The time resolution of the plot is one data point every 10 binary orbits.}
    \label{ae}
  \end{center}
\end{figure*}

Here we report the evolution of the binary eccentricity and the mass flow rates on to the sink particles in the simulations. In Figure~\ref{ae} we plot the semi--major axis and eccentricity evolution of the binary orbit with time for all of the simulations. In all plots the semi--major axis decreases with time, as expected from the capture of angular momentum and energy of retrograde gas orbits. The eccentricity evolution of the binary is consistent with the analytical predictions of \cite{Nixonetal2011a}. The circular binaries (for which $e \ll H/r$) remain circular as accretion at apocentre of the binary, which increases eccentricity, is offset by accretion at pericentre, which decreases eccentricity. For these simulations the eccentricity varies little and is approximately $10^{-4}$. The model  of \cite{Nixonetal2011a} predicts that the eccentricity $e$ should grow if it satisfies $e \gtrsim H/r$, and our simulations generally confirm this. Some of the simulations show eccentricity decay at early times before the disc has settled into a quasi-steady state. The only simulation which does not follow the predicted trend is $q=1.0$, $e=0.8$. For this case the eccentricity decays from $0.8$ to $0.799$ over the timescale of the simulation. It is unclear if the eccentricity would continue to decrease if the simulation were run for longer, or instead later increase.

\begin{figure*}
  \begin{center}
    \includegraphics[angle=0,width=0.5\textwidth]{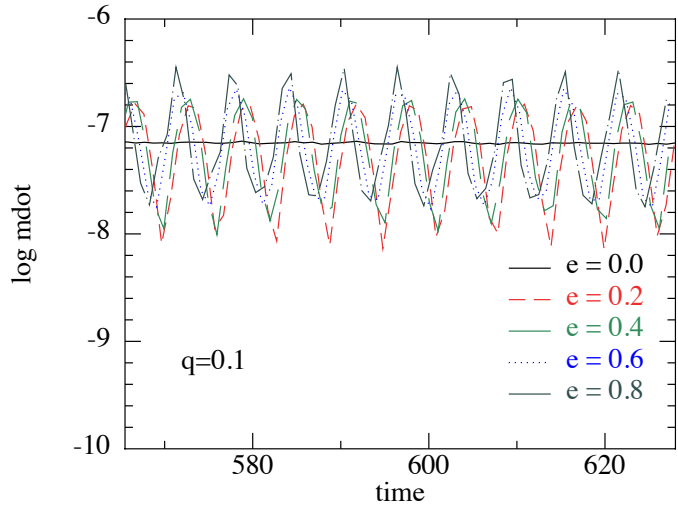}
    \caption{Accretion rate of gas flowing on to the binary for the $q=0.1$ simulations. This measures the mass flux through the sink particle radii. The accretion is binned with width equal to one binary dynamical time, $1/\Omega_{\rm b}$. The accretion is the total amount on to the primary and secondary sinks. Only the last 10 orbits of the binary are shown for clarity.}
    \label{mdot01}
  \end{center}
\end{figure*}

\begin{figure*}
  \begin{center}
    \includegraphics[angle=0,width=0.5\textwidth]{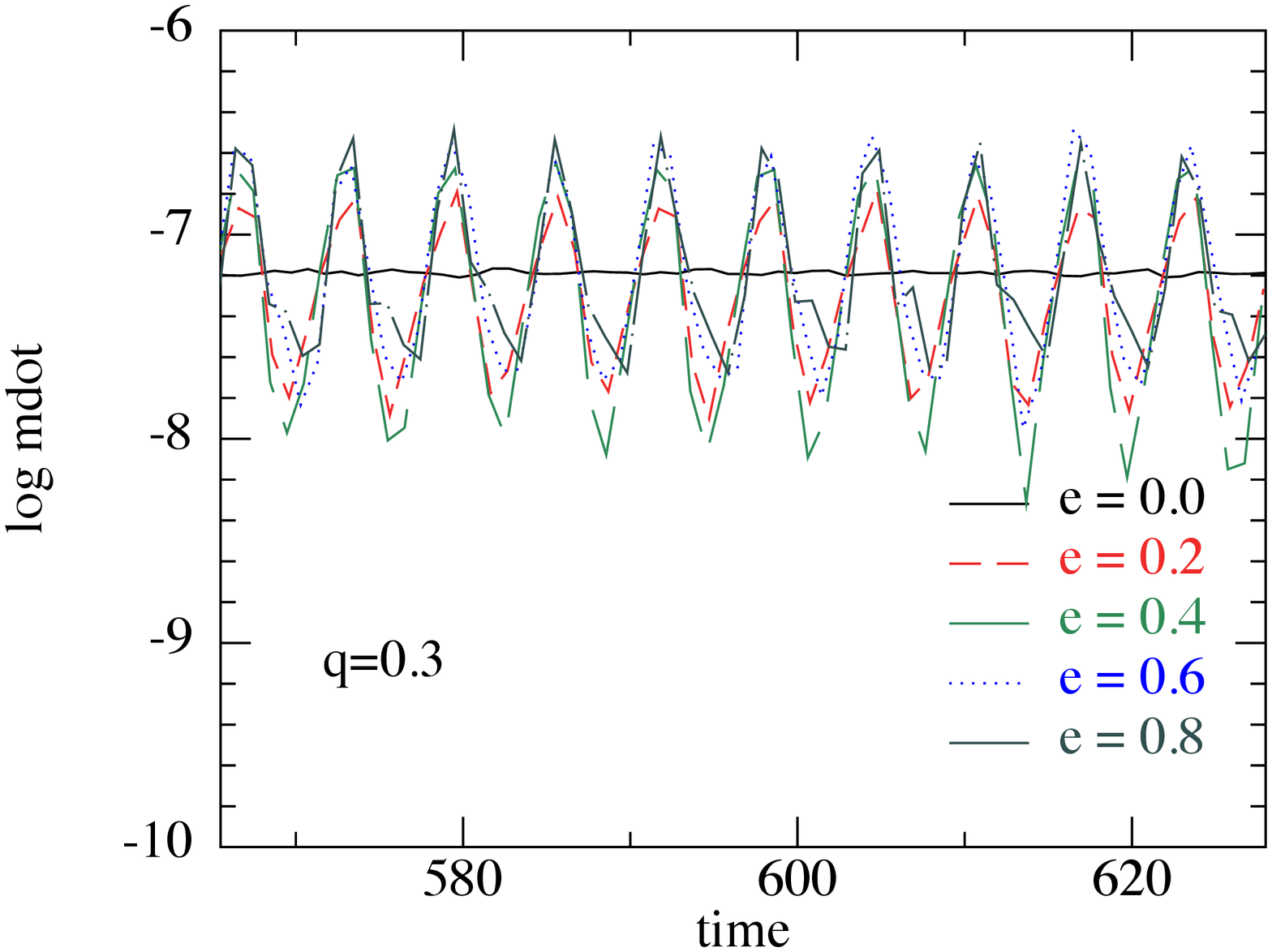}
    \caption{Same as Figure~\ref{mdot01} but for $q=0.3$.}
    \label{mdot03}
  \end{center}
\end{figure*}

\begin{figure*}
  \begin{center}
    \includegraphics[angle=0,width=0.5\textwidth]{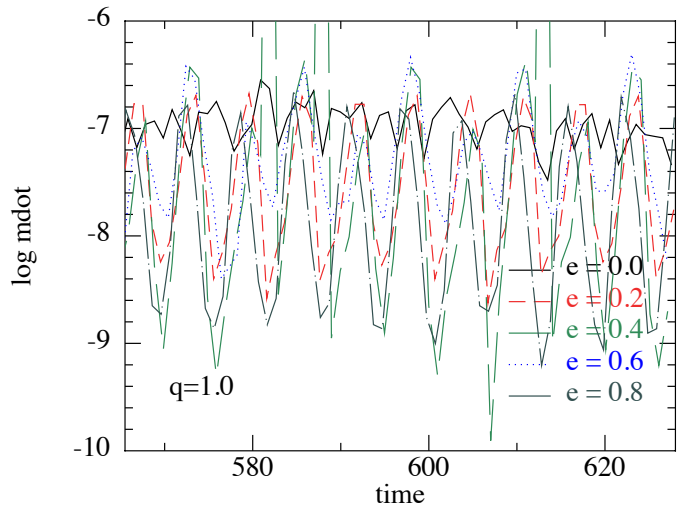}
    \caption{Same as Figure~\ref{mdot01} but for $q=1.0$.}
    \label{mdot10}
  \end{center}
\end{figure*}

The remaining eccentric simulations all show eccentricity growth with ${\rm d}e/{\rm d}t \sim 10^{-7}$ in units of time where $2\pi$ is one binary orbital period. This is in agreement with the analytical prediction of $\sim {\dot M}/M_2$ of \cite{Nixonetal2011a} (see Figures~\ref{mdot01}, \ref{mdot03}, and \ref{mdot10} for the accretion rates).  

In Figures~\ref{mdot01}, \ref{mdot03}, and \ref{mdot10}, we plot the accretion rate of gas on to the two sink particles as a function of time. The curves are binned with width equal to one binary dynamical time, $1/\Omega_{\rm b}$. We do not separate the primary and secondary accretion rates, as the secondary accretes almost all the gas, except for the equal mass case, where the rates are similar. In general the accretion rate is similar to the circular binary case, but with an obvious binary orbital period modulation. This modulation has an amplitude of approximately 1.5 orders of magnitude for $q=0.1$ and $q=0.3$, but up to $3-4$ orders of magnitude for the equal mass case. This periodicity is marked by strong accretion at apocentre and little accretion at pericentre of the binary orbit. In the prograde case, there is also a strong periodic modulation of accretion rate over the binary period for eccentric orbit binaries. However, the accretion rate
is highest near periastron \citep{AL1996}. 


\section{Discussion}
\label{sec:discussion}
We first discuss the limitations of our simulations, and possible improvements. We then consider the consequences of our results for possible astrophysical applications.

The simulations described above employ a simple locally isothermal thermodynamic treatment for the gas, where the sound speed is a radial power--law from the centre of mass. This approximation should not alter the fundamental behaviour of the discs, but does prevent any thermodynamic effects from arising. It would be interesting to run these simulations with an equation of state that accounts for shock and viscous heating while allowing the gas to cool on an appropriate timescale. This may affect the propagation of waves \citep[e.g.][]{LP1993} and thus the redistribution of angular momentum by the resonances.

Similarly our simulations neglect the effects of self-gravity in the gas, and the effects of magneto--hydrodynamic turbulence induced by the magneto--rotational instability, instead employing an accretion disc viscosity through an $\alpha$ parameter. These additional physical processes should be included in a full treatment of the problem, but are not expected to significantly alter our conclusions.

The simulations could also be run for longer to discover longer timescale evolution of the disc structures. However, the limiting factor is that accretion and disc spreading can deplete the gas to the point where there is too little of it in regions of interest. Therefore it would be worthwhile including mass input into the disc to allow a quasi--steady state to be achieved over a long timescale simulation. 

Binaries exist in a variety of astrophysical scenarios. However, to create a retrograde planar disc probably requires a chaotic environment which can introduce gas to the binary with random orientations (but see also e.g. \citealt{Pringle1996,Pringle1997} for a tilting instability which can cause the tilt of an initially planar disc to grow to retrograde angles). The two most likely chaotic scenarios are stellar binaries formed in dense star forming regions \citep[e.g.][]{Bateetal2010} and SMBH binaries accreting from the host galaxy \citep[e.g.][]{KP2006,KP2007,Kingetal2008}. In both cases, a significant fraction of randomly oriented discs are expected to align or counteralign with the binary plane depending primarily on the initial disc--binary inclination angle, but also on the ratio of disc to binary angular momentum \citep{Kingetal2005,Nixonetal2011b}. 

Retrograde accretion from circumbinary discs has recently been proposed by \cite{Nixonetal2011a} as a possible solution to the last parsec problem \citep{Begelmanetal1980,MM2001}. This model relied on the eccentricity growth, promoted by the capture of negative angular momentum gas, continuing to drive the binary eccentricities to large values, forcing the binary into the gravitational wave regime. If the resonances discussed here for retrograde circumbinary discs inhibit eccentricity growth as is expected in the prograde case for large eccentricities, then this may not be possible. However, these resonances require significant eccentricity to achieve large strengths, and even so are of significantly smaller strength than their prograde counterparts. Therefore it is quite possible that efficient eccentricity growth to values approaching unity is still possible. For this discussion the accretion rates on to the binary are highly suggestive, as the mean accretion rate appears to be insensitive to the binary eccentricity (see Figs \ref{mdot01}, \ref{mdot03}, and \ref{mdot10}). Instead the eccentricity acts to introduce significant periodicity in the mass flow rates through the sink accretion radii. As the binary is still capturing the same amount of material each orbit, it appears likely that the conclusions of \cite{Nixonetal2011a} are independent of the presence of these resonances. This also appears likely from Figure~\ref{ae} as the eccentricity generally grows in our simulations and at a rate consistent with the analytical prediction of \cite{Nixonetal2011a}. However, as discussed in Section \ref{sec:binev}, the assumptions of  the \cite{Nixonetal2011a} model may not be well satisfied at high eccentricity, We shall revisit this issue in more detail in future work.

\section{Conclusions}
\label{sec:conclusions}
We have presented an analytical model and three-dimensional hydrodynamical simulations of retrograde circumbinary discs covering a range in binary mass ratio and eccentricity. These simulations are consistent with the analytical prediction that binary eccentricity causes disc resonances to occur, which leads to angular momentum exchange between the binary and the disc. For circularly orbiting binaries, there are no disturbances (modes)  excited at Lindblad resonances in a retrograde disc. But for eccentric orbit binaries, Lindblad resonances can be excited in such a disc. The Lindblad torques increase with eccentricity. At high eccentricity $e \ga 0.6$, the most strongly excited disc mode $(-1,2)$  found in the simulations is also the mode expected to be most strongly excited in the analytic model  by  the Lindblad resonance at $\Omega(r) \simeq -\Omega_{\rm b}/3$. The Lindblad torques associated with this mode can open a gap in the disc at binary eccentricity $e \ga 0.6$ for typical disc parameters. The inner resonances are destabilized at high eccentricities. We also find that the binary semi--major axis and eccentricity evolution are in general agreement with the analytical predictions of \cite{Nixonetal2011a}.

The binary eccentricity evolution is affected by gas accretion and resonant interactions with the disc. Future simulations will focus on whether the binary eccentricity can grow to approximately unity or if there is a smaller limiting value. By including mass injection into the disc it will be possible to evolve the discs to a quasi-steady state and explore longer term evolution than is described here.

\section*{Acknowledgments}
We thank Andrew King for useful comments on the manuscript. CJN was supported for this work by NASA through the Einstein Fellowship Program, grant PF2--130098. S.H.L. acknowledges support from NASA grant NNX11AK61G. We used {\sc splash} \citep{Price2007} for the visualisation. This work utilised the Janus supercomputer, which is supported by the National Science Foundation (award number CNS-0821794) and the University of Colorado Boulder. The Janus supercomputer is a joint effort of the University of Colorado Boulder, the University of Colorado Denver and the National Center for Atmospheric Research. Janus is operated by the University of Colorado Boulder. 

\bibliographystyle{mn2e}
\bibliography{nixon}

\appendix

\section{Resolution}
\begin{figure}
  \begin{center}
    \includegraphics[angle=0,width=\textwidth]{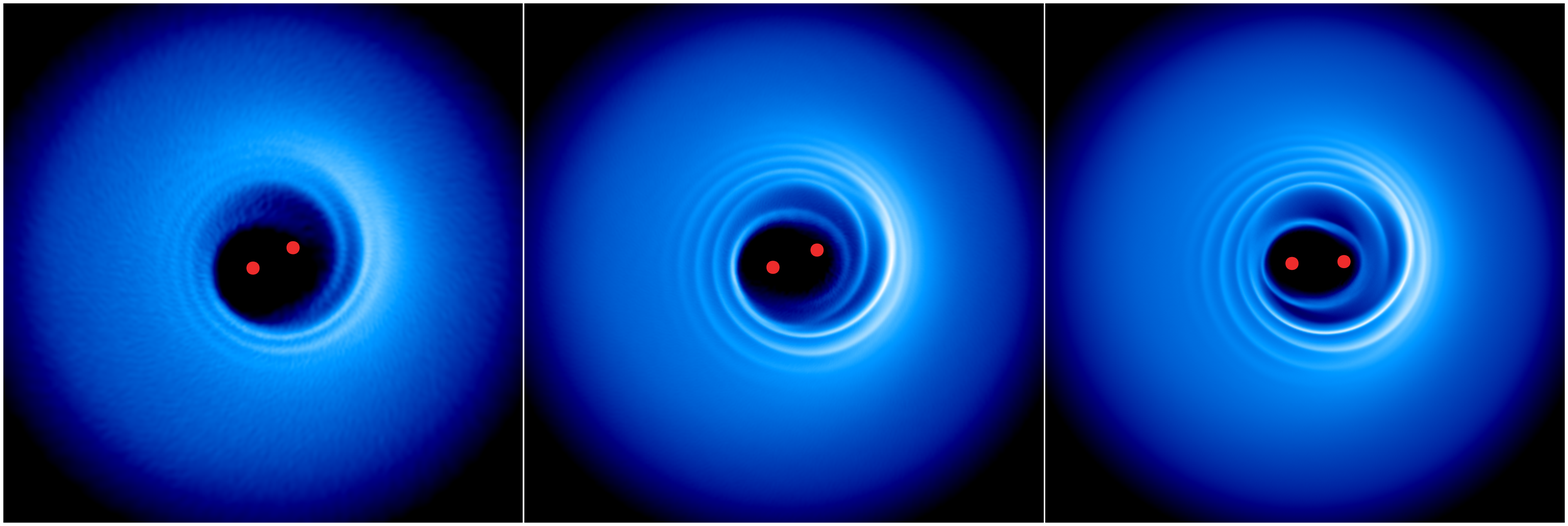}
    \caption{Surface  density rendering of the same simulation after 100 binary orbits, shown at three different resolutions. The colour scheme represents 2 orders of magnitude in surface density from the highest density (white) to the lowest density (black). The binary is represented by the two red filled circles, where their size denotes the size of their accretion radii. The three pictures are shown at precisely the same run time (100 binary orbits). The different phase of the binary orbit is due to the slightly different back reaction from the disc in each case.}
    \label{pic_comp}
  \end{center}
\end{figure}

\begin{figure}
  \begin{center}
    \includegraphics[angle=0,width=0.5\textwidth]{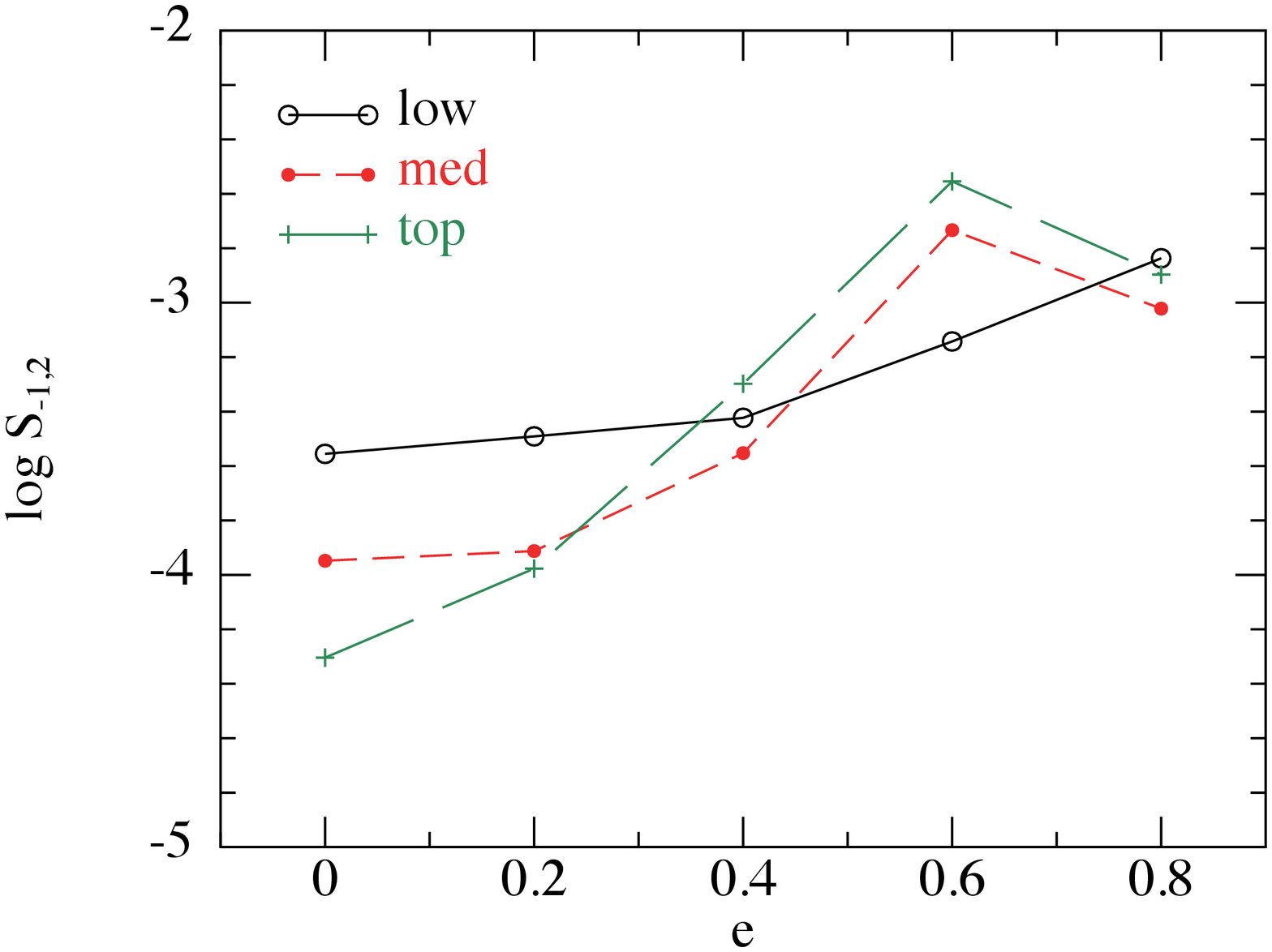}
    \caption{Comparison between low, medium and top resolution simulations for $q=0.3$. The black--solid line corresponds to the lowest resolution ($1.25\times 10^5$ particles), the red--dashed line corresponds to the medium resolution ($10^6$ particles) and the green--long-dashed line corresponds to the highest resolution ($8\times 10^6$ particles). For $e=0$ the mode strength should be zero and therefore the magnitude drops with increasing resolution. As the eccentricity is increased the mode strength increases. For the medium and highest resolution simulations good agreement is found, but at the lowest resolution it is clear the modes are not sufficiently resolved.}
    \label{rescomp}
  \end{center}
\end{figure}

Each simulation was run at three different resolutions corresponding to $1.25\times 10^{5}$, $10^6$ and $8\times 10^6$ particles. At each different resolution, the SPH viscosity was scaled so as to give the same physical viscosity $\alpha_{\rm SS} =0.05$ using the method described in Section 3.2.3 of \cite{LP2010}. In Figure~\ref{pic_comp} we show the surface  density rendering of the $q=0.3$, $e=0.6$ simulation at each resolution. It is clear from this figure that the lowest resolution simulation is under--resolved, but that at higher resolution the behaviour shows convergence. In Figure~\ref{rescomp} we show the $(-1,2)$ mode strength for each eccentricity at the different resolutions. This shows the same picture, where the lowest resolution simulations are not resolved, but the medium and highest resolution simulations show good agreement. For $e=0$ the mode strength drops by approximately the same factor as the resolution is increased -- this is to be expected as the mode strength should be formally zero. For higher eccentricities the difference between the medium and high resolution simulations is small.

\section{Application of Mode Strengths to SPH Simulations}
\label{sec:mssph}

In the application to SPH simulations, we consider  $N$ particles each of unit mass that are located at  the 3D Cartesian positions $(x_j(\tau), y_j(\tau),  z_j(\tau))$ and approximate the surface density distribution as
\begin{equation}
\label{sigmadelta}
\Sigma(x, y, \tau) = \sum \limits_{j=1}^N \delta(x-x_j(\tau)) \delta(y-y_j(\tau))
\end{equation}
for a disc of mass $N$. In general, we have
\begin{eqnarray}
\tilde{\Sigma}_{\rm{f, g}, \ell, m} &=&  \int_0^{\infty} \Sigma_{\rm{f, g}, \ell, m}(r) \,r \, dr \\
   &=&  \frac{1}{2 \pi^2 (1 + \delta_{ \ell,0} \delta_{m,0}) } \int_0^{2 \pi} \int_0^{2 \pi}  \int_0^{\infty} \Sigma(r, \theta, \tau)  f(m \theta) g(\ell \tau)  \, \,r \, dr \,  d\theta \, d\tau \\
   &=&  \frac{1}{2 \pi^2  (1 + \delta_{ \ell,0} \delta_{m,0})} \int_0^{2 \pi} \int_{-\infty}^{\infty}   \int_{-\infty}^{\infty} \Sigma(x,y, \tau) 
   f(m \, \theta(x,y)) g(\ell \,\tau)  \,  dx \, dy \, d\tau \\
   &=&   \frac{1}{2 \pi^2 (1 + \delta_{ \ell,0} \delta_{m,0})} \int_0^{2 \pi}  \sum \limits_{j=1}^N f{( m \, \theta_j(\tau))} g(\ell \,\tau)  \, d\tau, 
\label{Siglmsph}
\end{eqnarray}
where  $f$ and $g$ each can be the $\cos$ or $\sin$ functions and $\theta_j(\tau) = \arctan{(y_j(\tau)/x_j(\tau))}$.

As a illustration of this analysis, we consider a set of $N$ particles labelled by $j$ at radius $r=1$ having angular location 
\begin{equation}
\theta_j = \psi_j + \lambda \cos{(m \psi_j - \ell \tau)},
\label{sphtestpos}
\end{equation}
where
\begin{equation}
\psi_j = \frac{2 \pi \, j}{N} + \tau
\end{equation}
for integer $1 \le j \le N$ and $|\lambda| \ll 1$. It then follows that the density distribution of particles is
\begin{eqnarray}
\Sigma(r, \psi_j, t) &\simeq& \delta(r-1)/(d \theta_j/dj)\\
            &=& \delta(r-1) (\frac{2 \pi}{N}(1 - m \lambda \sin{(m \psi_j - \ell \tau)}))^{-1} \\
            & \simeq& \frac{N}{2 \pi} \delta(r-1) (1 + m \lambda \sin{(m \psi_j - \ell \tau)}).
 \label{Sigexample}
\end{eqnarray}
For $\ell \ne 0$ or $m \ne 0$, we have
\begin{eqnarray}
\tilde{\Sigma}_{\ell, m} &=& -\frac{i m \, N \lambda}{2 \pi},
\label{Siglmex}
\end{eqnarray}
where the $-i$ factor is included to obtain the $\sin({m \psi_j - \ell \tau})$ function in Equation (\ref{Sigexample}). 

We show below that this expression for $\tilde{\Sigma}_{\ell, m}$ can be also obtained from the use of Equations  (\ref{Siglmreim}) and (\ref{Siglmsph}). We use the fact that
\begin{eqnarray}
\sin{(m \theta_j)} & = & \sin{(m \psi_j + m \lambda  \cos{(m \psi_j - \ell \tau)})},\\
     & \simeq & \sin{(m \psi_j)} + m  \lambda [\cos^2{(m \psi_j)} \cos{(\ell \tau)} + \cos{(m \psi_j)} \sin{(m \psi_j)} \sin{(\ell \tau) } ],
\end{eqnarray}
since $|\lambda| \ll 1$. We then evaluate 
\begin{eqnarray}
\tilde{\Sigma}_{\rm{cos, sin}, \ell, m}  &=& \frac{1}{2 \pi^2} \int_0^{2 \pi} \sum \limits_{j=1}^N \sin{( m \, \theta_j(\tau))} \cos(\ell \,\tau) \, d\tau \\
&\simeq& \frac{1}{2 \pi^2} \int_0^{2 \pi} \sum \limits_{j=1}^N  [\sin{(m \psi_j)}  \cos{(\ell \,\tau)} + m  \lambda \cos^2{(m \psi_j)} \cos^2{(\ell \tau)}  + \nonumber \\ 
& &  \hspace{1in} m  \lambda \cos{(m \psi_j)} \sin{(m \psi_j)} \sin{(\ell \tau) }  \cos{(\ell \,\tau)})] \, d\tau \\
&=& \frac{m  \lambda }{2 \pi^2} \int_0^{2 \pi} \sum \limits_{j=1}^N  \cos^2{(m \psi_j)} \cos^2(\ell \,\tau) \, d\tau \\
&=& \frac{m  N \lambda }{4 \pi}.
\end{eqnarray}
Similarly, it can be shown that
\begin{equation}
\tilde{\Sigma}_{\rm{sin, cos}, \ell, m} = -\frac{m  N \lambda }{4 \pi},
\end{equation}
\begin{equation}
\tilde{\Sigma}_{\rm{cos, cos}, \ell, m} = 0,
\end{equation}
\begin{equation}
\tilde{\Sigma}_{\rm{sin, sin}, \ell, m} = 0.
\end{equation}
From equation (\ref{Siglmreim}), we have
\begin{eqnarray}
\tilde{\Sigma}_{\ell, m} &=&  
 \, i(\tilde{\Sigma}_{\rm{sin, cos}, \ell, m}  -  \tilde{\Sigma}_{\rm{cos, sin}, \ell, m}), \\
 &=& -\frac{i m  N \lambda }{2 \pi},
 \label{Siglmex1}
\end{eqnarray}
that agrees with Equation (\ref{Siglmex}) and so analytically shows the validity of the decomposition prescription. We also have that
\begin{eqnarray}
S_{\ell,m} &=& \frac{2 \pi |\tilde{\Sigma}_{\ell, m}|}{N}\\
&=& \frac{2 \pi m N |\lambda|}{2 \pi N} \\
&=& m |\lambda|.
\label{Slam}
\end{eqnarray}

\end{document}